\documentclass[twocolumn,showpacs,preprintnumbers,amsmath,amssymb]{revtex4}

\usepackage{graphicx}
\usepackage{dcolumn}
\usepackage{bm}

\newcommand{\be}{\begin{equation}}
\newcommand{\ee}{\end{equation}}
\newcommand{\bea}{\setlength \arraycolsep{1pt} \begin{eqnarray}}
\newcommand{\eea}{\end{eqnarray}}

\begin{document}

\title{Solvent Entropy in and Coarse-Graining of Polymer Lattice Models}

\author{Pengfei Zhang and Qiang Wang\email{q.wang@colostate.edu}}
\affiliation{Department of Chemical and Biological Engineering, Colorado State University, Fort Collins, CO 80523-1370}

\date{\today}

\begin{abstract}
The self- and mutual-avoiding walk used in conventional lattice models for polymeric systems requires that all lattice sites, polymer segments, and solvent molecules (unoccupied lattice sites) have the same volume. This incorrectly accounts for the solvent entropy (i.e., size ratio between polymer segments and solvent molecules), and also limits the coarse-graining capability of such models, where the invariant degree of polymerization controlling the system fluctuations is too small (thus exaggerating the system fluctuations) compared to that in most experiments. Here we show how to properly account for the solvent entropy in the recently proposed lattice models with multiple occupancy of lattice sites [Q. Wang, Soft Matter \textbf{5}, 4564 (2009)], and present a quantitative coarse-graining strategy that ensures both the solvent entropy and the fluctuations in the original systems are properly accounted for using such lattice models. Although proposed based on homogeneous polymer solutions, our strategy is equally applicable to inhomogeneous systems such as polymer brushes immersed in a small-molecule solvent.
\end{abstract}

\maketitle

Lattice models, where the continuum space is divided into a lattice, are ubiquitously used in both theoretical and simulation studies of polymers; perhaps the most well-known example is in the original Flory-Huggins theory\cite{FH}. In such conventional lattice models, each lattice site is occupied by at most one polymer segment representing a group of real monomers, and an unoccupied lattice site is often treated as a solvent molecule; in accordance with this self- and mutual-avoiding walk (SMAW), the nearest-neighbor interactions are used. We note that Monte Carlo (MC) simulations using conventional lattice models are already much faster than off-lattice simulations by avoiding the time-consuming pair-potential evaluation in the latter, which leads to their wide applications.

There are, however, some disadvantages of using a lattice model. For example, a lattice is anisotropic and may cause some artifacts. While this problem cannot be completely eliminated, one of us recently showed how to quantify and thus minimize the lattice anisotropy.\cite{AniLat} More significantly, in conventional lattice models, SMAW requires that all lattice sites, polymer segments, and solvent molecules have the same volume. Since a polymer segment here is the coarse-grained representation of a group of real monomers, this incorrectly accounts for the solvent entropy (i.e., the size ratio between polymer segments and solvent molecules). It also limits the coarse-graining capability of such models. For example, Binder and co-workers showed that one bond in the bond-fluctuation model\cite{BFM} maps to five consecutive united-atom bonds of polyethylene;\cite{BFMPE} chains of $N \sim 10^2$ segments in this lattice model, which represents in most cases the upper-limit of $N$ in many-chain simulations with hard excluded-volume interactions (e.g., SMAW), are still too short compared to those used in experiments.

The fluctuations in polymeric systems are mainly controlled by the invariant degree of polymerization $\bar{\mathcal{N}} \equiv \left(n R_e^3 / V\right)^2$, where $n$ denotes the number of chains in volume $V$ and $R_e$ the root-mean-square chain end-to-end distance. Essentially, $\bar{\mathcal{N}}$ quantifies how many chains a single chain interacts with within its volume; when $\bar{\mathcal{N}}$ is large, the system fluctuations are small and the mean-field approximation neglecting fluctations/correlations becomes more accurate. Taking polymer melts as an example, we have $R_e = \sqrt{N-1} b$, where $b$ denotes the root-mean-square bond length of an ideal chain, and with hard excluded-volume interactions $V \sim n N b^3$; this means $\bar{\mathcal{N}}$ is on the same order of magnitude as $N$. Since $\bar{\mathcal{N}}$ is at least $10^3$ in most experiments, \emph{conventional molecular simulations with hard excluded-volume interactions (thus small $\bar{\mathcal{N}}$- or $N$-values) exaggerate the fluctuations in many-chain systems} such as concentrated polymer solutions or melts.

Recently, one of us proposed fast lattice Monte Carlo (FLMC) simulation, where multiple occupancy of lattice sites (MOLS) is allowed with a proper Boltzmann weight.\cite{FLMC} The use of soft potentials that allow particle overlapping (e.g., MOLS), as in its off-lattice analog\cite{FOMC}, greatly speeds-up the sampling of configuration space. More significantly, $V \sim n N b^3$ (for polymer melts) no longer holds, and $\bar{\mathcal{N}}$ and $N$ are thus decoupled; in such simulations one can therefore study systems at much larger $\bar{\mathcal{N}}$ (e.g., with experimentally accessible fluctuations) while $N$ becomes a chain discretization (coarse-graining) parameter that does not correspond to the actual chain length used in experiments. This point is crucial for understanding coarse-grained models with soft potentials. MOLS further allows the use of Kronecker $\delta$-function interactions (which are isotropic on any lattice) instead of the nearest-neighbor interactions (which are anisotropic except in 1D).\cite{FLMC} As demonstrated in Ref.~\cite{FLMC}, FLMC simulations (with MOLS and Kronecker $\delta$-function interactions) are much more efficient, in the study of equilibrium properties of soft matter such as polymers, than both conventional lattice MC simulations (with SMAW and nearest-neighbor interactions) and fast off-lattice MC simulations (with pair-potential calculations)\cite{FOMC}. Last but not least, when compared with the corresponding lattice field theories based on the \emph{same} Hamiltonian, FLMC simulations provide a powerful means for unambiguously and quantitatively revealing the effects of long-wavelength correlations/fluctuations in the system, as demonstrated in our recent work\cite{FLMC,1DBulk,BrushImp,BrushExp}.

In this Letter, we show how to properly account for the solvent entropy in lattice models with MOLS, and how to use such models as quantitative, coarse-grained description of polymeric systems, thus overcoming the aforementioned drawback of conventional lattice models.

Let us consider, as the original system for coarse-graining, a homogeneous and incompressible mixture of $n$ homopolymer chains each of $N_m$ monomers and $n_{\rm S}$ solvent (S) molecules. With $v$ and $v_{\rm S}$ denoting the volume of each monomer and solvent molecule, respectively, the average volume fraction of polymers in the mixture is given by $\bar{\phi} = n N_m v / V$, where $V = n N_m v + n_{\rm S} v_{\rm S}$. In the coarse-grained model, we have $n$ chains each of $N$ segments and $n_{\rm S}$ solvent molecules on a lattice. Each polymer segment or solvent molecule occupies one lattice site, and each lattice site is occupied by any combination of $i \ge 0$ polymer segments and $j \ge 0$ solvent molecules such that $i + j/r = \rho_0$ (note that $r>0$ and $\rho_0>0$ here do not have to be integers). The average volume fraction of polymers in the coarse-grained model is then given by $\bar{\phi} = n N l^3 / \rho_0 V$, where the lattice volume $V = (n N / \rho_0 + n_{\rm S} / r \rho_0) l^3$ and $l$ is the lattice spacing.

To quantitatively map the original system to the coarse-grained model, the values of $r$, $\rho_0$, and $l$ need to be determined. $r$ represents the size ratio between a polymer segment and a solvent molecule, and setting $r = N_m v / N v_{\rm S}$ gives the same $\bar{\phi}$ for the original and the coarse-grained systems. Setting $\rho_0 = N l^3 / N_m v$ further gives the same $V$. With this mapping, it is easy to show that our non-bonded interaction energy $\mathcal{H}^E$ in the coarse-grained model (given by Eq.~(\ref{eq:HE}) below) is also the same as that in the original system given by $\beta \mathcal{H}^E = \chi \bar{\phi} (1 - \bar{\phi}) V / v_{\rm S}$\cite{chi}, where $\beta \equiv 1 / k_B T$ with $k_B$ being the Boltzmann constant and $T$ the thermodynamic temperature, and $\chi$ is the Flory-Huggins interaction parameter between a monomer and a solvent molecule. Finally, to obtain the same $\bar{\mathcal{N}}$, we set $R_e$ to be the same, i.e.,
\be \label{eq:Re}
R_{e,\rm O}(N_m,\chi,\bar{\phi}) = R_{e,\rm CG}(N,\chi,\bar{\phi},r,\rho_0,b),
\ee
where the subscripts ``O'' and ``CG'' denote, respectively, the original and the coarse-grained systems. Note that $b$ is only a function of $l$, which depends on the lattice type: for example, $b=l$ for the simple cubic lattice (SCL) and $\sqrt{2} l$ for the face-centered cubic lattice; for a chosen lattice, $l$ can therefore be solved from Eq.~(\ref{eq:Re}). Under the assumption of ideal chain conformations good for polymer melts ($\bar{\phi}=1$) or solutions in a $\theta$-solvent ($\chi=0.5$), Eq.~(\ref{eq:Re}) gives
\be \label{eq:l}
l = \sqrt{(N_m-1) / (N-1)} a (b/l)^{-1}
\ee
with $a$ being the polymer statistical segment length.

This simple yet quantitative coarse-graining strategy for lattice models with MOLS is the central result of this Letter. It ensures that both the solvent entropy and fluctuations in the original system are properly accounted for in the coarse-grained model, and is equally applicable to inhomogeneous polymeric systems. To demonstrate this, we apply it to inhomogeneous systems of polymer brush immersed in a small-molecule solvent.

A polymer brush refers to chains end-grafted on a substrate at high densities such that the excluded-volume interactions among adjacent chains make them stretch perpendicular to the substrate.\cite{Brush} Here we consider the simplest case of uncompressed homopolymer brush on a flat substrate, which has been widely studied by experiments, molecular simulations, and various theories.\cite{BrushImp,BrushExp,Brush,Rev} In particular, neutron-reflectivity (NR) studies by several groups have measured the polymer volume fraction profile in the direction perpendicular to the substrate, denoted by $\phi(x)$. It was such studies\cite{NR} in early 1990s that unambiguously confirmed the parabolic profile of $\phi(x)$ for brushes in a good solvent as predicted by the infinite-stretching limit (ISL) theory of Milner et~al.\cite{Miln}, instead of the step-function-like profile assumed in the scaling theory of Alexander and de Gennes\cite{AxdG}. Table~\ref{Exp} lists the brush systems studied in some experiments, where the dimensionless chain-grafting density $\sigma$ is defined as the number of chains per substrate area of $R_e^2$, and Table~\ref{Den} lists the polymer and solvent densities used in our modeling. In addition, we use $a=0.67$~nm for polystyrene (PS)\cite{PSa} regardless of deuteration.
\begin{table}[t]
\caption{\label{Exp} Brush systems studied in some neutron-reflectivity experiments.}
\begin{ruledtabular}
\begin{tabular}{ccccc}
Ref. & Brush & Solvent & $N_m$ & $\sigma$ \\
\hline
\cite{Kent} & d-PS$^a$ & dioctyl phthalate & 1515 & 12.76 \\
\cite{Karim} & PS & d-toluene & 1008 & 27.19 \\
\end{tabular}
\end{ruledtabular}
\begin{flushleft}
$^a$~``d'' denotes deuteration, and PS refers to polystyrene.
\end{flushleft}
\end{table}
\begin{table}[t]
\caption{\label{Den} Polymer and solvent densities.}
\begin{ruledtabular}
\begin{tabular}{ccccc}
Polymer/Solvent & Density (g/cm$^3$) & Ref. \\
\hline
PS & 1.047 & \cite{PS_rho} \\
d-PS & 1.12 & \cite{dPS_rho} \\
dioctyl phthalate & 0.984 & \cite{dPH_rho} \\
d-toluene & 0.943 & \cite{dPS_rho} \\
\end{tabular}
\end{ruledtabular}
\end{table}

In our coarse-grained lattice model, the first segment of all chains is grafted at $x=l$ (i.e., in the first lattice layer), and an impenetrable substrate is placed at $x=0$, which cannot be occupied by polymer segments or solvent molecules. There are totally $N$ lattice layers in the $x$-direction. The canonical-ensemble configuration integral of this model system is
\bea \label{eq:Z}
\mathcal{Z} & = & \prod_{k=1}^n \prod_{s=1}^N \sum_{{\bf R}_{k,s}} \cdot \prod_{k=1}^{n_\mathrm{S}} \sum_{{\bf r}_k} \cdot \exp \left(-\beta \sum_{k=1}^n h_k^C - \beta \mathcal{H}^E \right) \nonumber \\
& & \cdot \prod_{\bf r} \delta_{\hat{\rho}({\bf r}) + \hat{\rho}_\mathrm{S}({\bf r})/r,\rho_0},
\eea
where ${\bf R}_{k,s}$ denotes the lattice position of the $s^{\mathrm {th}}$ segment on the $k^{\mathrm {th}}$ chain, ${\bf r}_k$ denotes the position of the $k^{\mathrm{th}}$ solvent molecule, the summations are over all possible positions of all polymer segments and solvent molecules, respectively, and ``$\cdot$'' means that the products before it do not apply to the terms after it but the summations before it do. Furthermore, $h_k^C$ is the Hamiltonian of the $k^{\rm th}$ chain due to its connectivity; for single-bond lattice models (e.g., SCL), $\beta h_k^C=0$ if the chain connectivity is maintained and $\infty$ otherwise. The Hamiltonian $\mathcal{H}^E$ due to non-bonded interactions is
\be \label{eq:HE}
\beta \mathcal{H}^E\left[\hat{\rho},\hat{\rho}_\mathrm{S}\right] = \frac{\chi}{\rho_0} \sum_{{\bf r}} \hat{\rho}({\bf r}) \hat{\rho}_{\mathrm{S}}({\bf r}),
\ee
where $\hat{\rho}({\bf r}) \equiv \sum_{k=1}^{n}\sum_{s=1}^{N} \delta_{{\bf r},{\bf R}_{k,s}}$ and $\hat{\rho}_{\mathrm{S}}({\bf r}) \equiv \sum_{k=1}^{n_{\mathrm{S}}} \delta_{{\bf r},{\bf r}_k}$ are the microscopic number density of polymer segments and solvent molecules, respectively, at lattice site ${\bf r}$. Finally, $\delta$ denotes the Kronecker $\delta$-function.

Eqs.~(\ref{eq:Z}) and (\ref{eq:HE}) are similar to those in our previous work\cite{BrushExp}, where brushes in 1D with $r=1$ were studied. Following Ref.~\cite{BrushExp}, one can derive the corresponding lattice self-consistent field (LSCF) theory, which is a mean-field theory neglecting the system fluctuations/correlations and is exact in the limit of $\rho_0 \to \infty$ (i.e., $n \to \infty$, or equivalently $\sigma \to \infty$). With uniform grafting (i.e., all lattice sites at $x=l$ are grafted with the same number of chains), LSCF equations for brushes in a good or $\theta$-solvent need to be solved only in the $x$-direction, with five parameters determining $\phi(x)$: $N$, $\sigma/\rho_0 = \bar{\phi} (R_e/l)^2$, $r$, $\chi$, and $\lambda_0$; the last one, defined as the fraction of the nearest-neighbor lattice sites that are at the same $x$ of a given site, is a property of the chosen lattice.

\begin{figure}[t]
\begin{center}
\includegraphics[height=5.4cm]{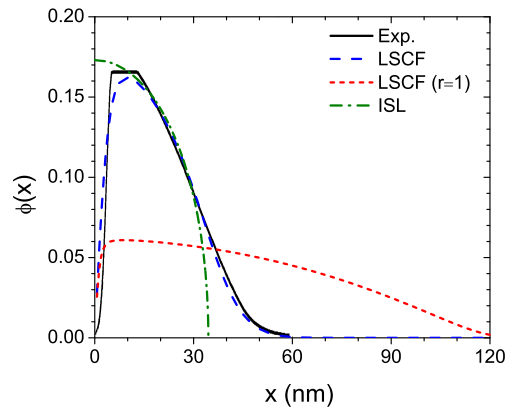}
\end{center}
\caption{\label{Kent}(Color online) Comparisons of $\phi(x)$ obtained from the NR measurements by Kent et~al.\cite{Kent} and our LSCF and ISL calculations with the appropriate $r$-value; $\phi(x)$ obtained from LSCF calculations with $r=1$ as in conventional lattice models is also shown. $N=N_m$, $\chi=0.5$, and $\lambda_0 = 2/3$.}
\end{figure}

We first perform LSCF calculations to examine the effects of solvent entropy (i.e., $r$), where we use $N=N_m$ and $\lambda_0 = 2/3$ corresponding to SCL. Kent et~al.\ studied d-PS brushes in a $\theta$-solvent, dioctyl phthalate, where $v/v_{\rm S} \approx 0.252$.\cite{Kent} Fig.~\ref{Kent} shows nearly quantitative agreement on $\phi(x)$ between their NR measurements and our LSCF calculations using the appropriate $r$, without any parameter-fitting; similar agreement is also obtained for other $\sigma$-values studied in Ref.~\cite{Kent} (data not shown). We also see that $r=1$, as used in conventional lattice models, largely overestimates the solvent entropy in this case, thus leading to much flatter $\phi(x)$.
\begin{figure}[t]
\begin{center}
\includegraphics[height=5.4cm]{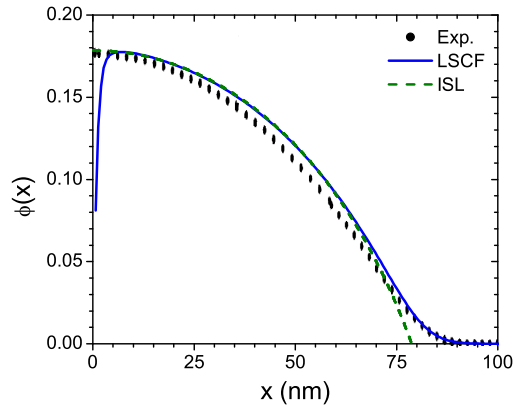}
\end{center}
\caption{\label{Karim}(Color online) Comparisons of $\phi(x)$ obtained from the NR measurements by Karim et~al.\cite{Karim} and our LSCF and ISL calculations with the appropriate $r$-value. $N=N_m$, $\chi=0.4$, and $\lambda_0 = 2/3$.}
\end{figure}

Fig.~\ref{Karim} shows good agreement on $\phi(x)$ between the NR measurements of Karim et~al.\ on PS brush in a good solvent, d-toluene, where $v/v_{\rm S} \approx 0.936$,\cite{Karim} and our LSCF calculations with $\chi = 0.4$. The largest discrepancy is in the depletion zone near the substrate, which was not found in experiments probably due to the attraction between PS and the silicon substrate\cite{Karim}. We note that, while conventional lattice models with $N=N_m$ can closely capture the solvent entropy in this system, the appropriate $r$-value must be used in coarse-grained models with $N \ll N_m$.

\begin{figure}[t]
\begin{center}
\includegraphics[height=5.4cm]{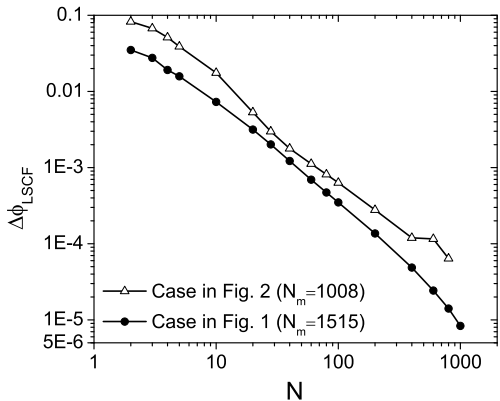}
\end{center}
\caption{\label{CGexp} Deviation of $\phi(x)$ obtained from LSCF calculations with $N$ from that with $N_m$, $\Delta\phi_{\rm LSCF}$, for the cases shown in Figs.~\ref{Kent} and \ref{Karim}.}
\end{figure}

Next, we use LSCF calculations on SCL to examine the effects of coarse-graining (i.e., $N$). Fig.~\ref{CGexp} shows the deviation of $\phi(x)$ obtained with $N$ from that with $N_m$, $\Delta\phi_{\rm LSCF} \equiv \sqrt{\int_0^{N l(N)} {\rm d}x \left[\phi_{\rm LSCF}(x;N) - \phi_{\rm LSCF}(x;N_m)\right]^2 \Big/ N l(N)}$, vs.\ $N$ for the two cases shown in Figs.~\ref{Kent} and \ref{Karim}. Clearly, coarse-grained models cannot exactly reproduce the original system in all aspects. But we see only small deviations within large degrees of coarse-graining, e.g., $\Delta\phi_{\rm LSCF} < 0.01$ for $N \gtrsim 20$, indicating the success of our quantitative coarse-graining strategy for lattice models with MOLS.

Some discussion on the quantitative comparisons with NR measurements is in order here. For simplicity, we have used Eq.~(\ref{eq:l}) instead of Eq.~(\ref{eq:Re}) in the above results. While this is less accurate for PS in d-toluene, it could be offset by adjusting the $\chi$-value. The unquantified polymer-substrate interaction and the fitting procedure used to obtain $\phi(x)$ in NR measurements also preclude unambiguous comparisons with our results using a simplified lattice model.

\begin{figure}[t]
\begin{center}
\includegraphics[height=5.4cm]{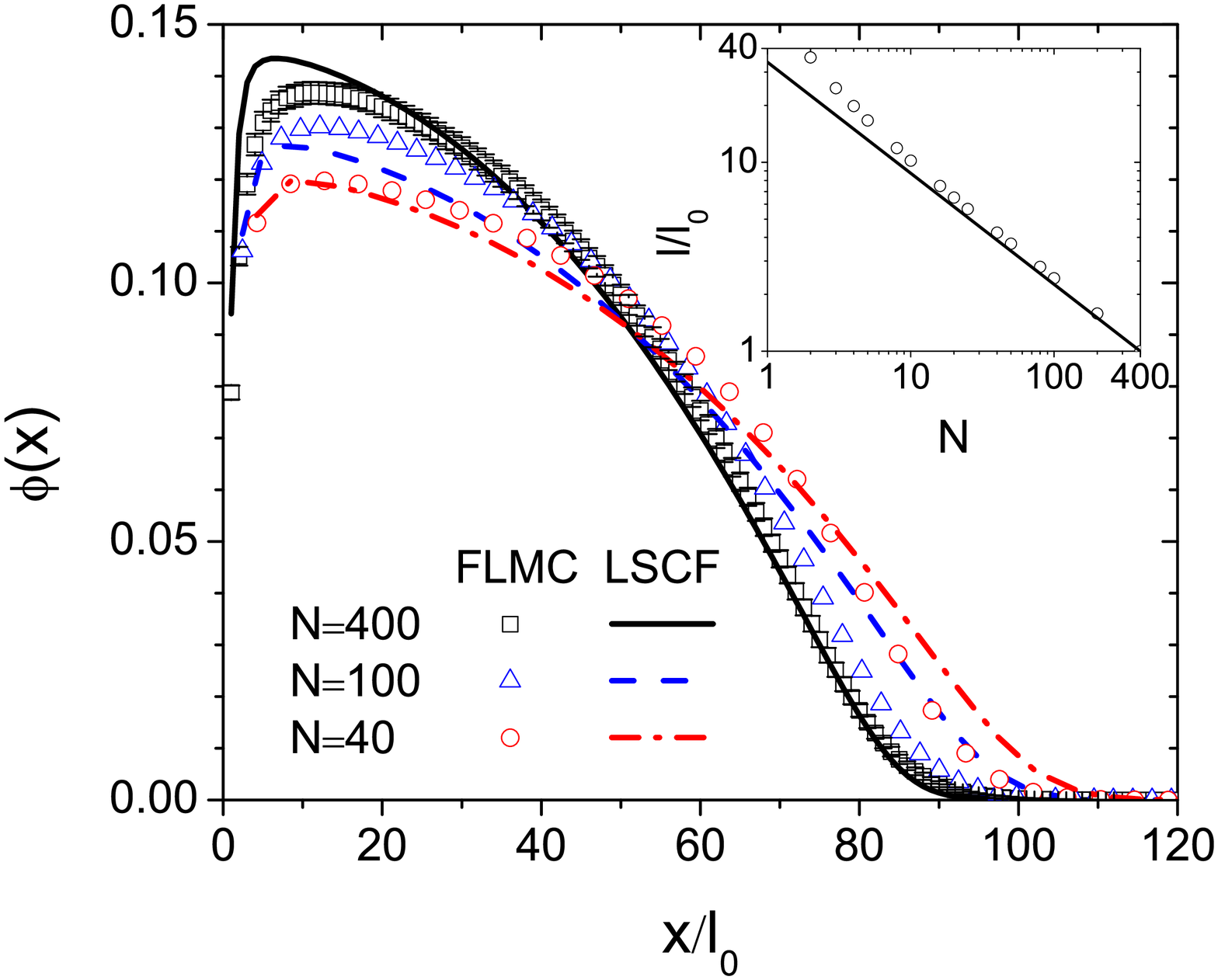}
\end{center}
\caption{\label{Kara}(Color online) Comparisons of $\phi(x)$ obtained from FLMC simulations and 1D LSCF calculations with the appropriate $r$-value. The error bars in $\phi_{\rm FLMC}(x)$ with $N=100$ and 40 are smaller than the symbols and thus not shown. $N_m=400$, $\chi=0$, and $\lambda_0 = 2/3$. The inset shows how $l$ obtained from Eq.~(\ref{eq:Re}) varies with $N$, where the straight line has a slope of $-0.5876$ as given by the limit of an infinitely long chain of self-avoiding walk\cite{SAWnu}.}
\end{figure}

To avoid such ambiguities and to quantitatively reveal the system fluctuations/correlations, we now choose an athermal ($\chi = 0$) system of SMAW chains with $N_m=400$ and $\sigma = 0.02 R_e^2$ (thus $\bar{\phi}=0.02$) on SCL with an impenetrable grafting substrate as the original system. We have $R_{e,\rm O} = (35.86 \pm 0.12) l_0$ from bulk MC simulations, where $l_0$ denotes the lattice spacing of the original system, and solve Eq.~(\ref{eq:Re}) for $l$ at various $N$ with $R_{e,\rm CG}(N, r=N_m/N, \rho_0 = N l^3 / N_m l_0^3, b=l)$ obtained via bulk MC simulations on SCL. The inset of Fig.~\ref{Kara} shows that $l$ monotonically increases with decreasing $N$.

For $N=N_m$ (thus $l=l_0$), Fig.~\ref{Kara} compares $\phi(x)$ obtained from MC simulations (where chains are grafted as uniformly as possible) and 1D LSCF calculations of the original system; the difference is due to the system fluctuations/correlations neglected by the LSCF theory\cite{Note1}, which make the chains more stretched (i.e., $\phi_{\rm FLMC}(x)$ is flatter than $\phi_{\rm LSCF}(x)$) as found in our previous work on brushes in an implicit, good solvent\cite{BrushImp} and 1D brushes in an explicit solvent with $r=1$\cite{BrushExp}. The opposite, however, occurs for $r=4$ and 10 on SCL, as shown also in the figure.

\begin{figure}[t]
\begin{center}
\includegraphics[height=5.4cm]{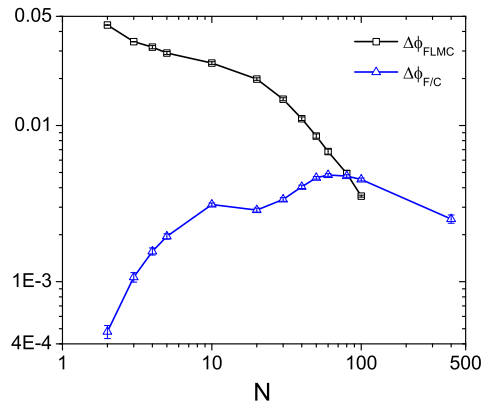}
\end{center}
\caption{\label{CGMC}(Color online) Deviation of $\phi(x)$ obtained from FLMC simulations with $N$ from that with $N_m=400$, $\Delta\phi_{\rm FLMC}$, for the case shown in Fig.~\ref{Kara}.}
\end{figure}
To quantify the effects of coarse-graining on $\phi(x)$, Fig.~\ref{CGMC} shows $\Delta\phi_{\rm FLMC} \equiv \sqrt{\int_0^{N l(N)} {\rm d}x \left[\phi_{\rm FLMC}(x;N) - \phi_{\rm FLMC}(x;N_m)\right]^2 \Big/ N l(N)}$, vs.\ $N$; we again see only small deviations within large degrees of coarse-graining, e.g., $\Delta\phi_{\rm FLMC} < 0.01$ for $N \gtrsim 50$. We can further quantify the effects of coarse-graining on the system fluctuations/correlations, measured by $\Delta\phi_{\rm F/C} \equiv \sqrt{\int_0^{N l(N)} {\rm d}x \left[\phi_{\rm LSCF}(x;N) - \phi_{\rm FLMC}(x;N)\right]^2 \Big/ N l(N)}$. Fig.~\ref{CGMC} shows that, while $\Delta\phi_{\rm F/C}$ exhibits small variation with $N$ as expected for inhomogeneous systems, our coarse-graining strategy successfully preserves its order of magnitude ($10^{-3}$) for $N \in [3, N_m]$; note that, for small $N$ ($\lesssim 10$), $\Delta\phi_{\rm F/C}$ is expected to decrease with decreasing $N$ and vanish when $N=1$.

Finally, we note that the correct solvent entropy can be readily incorporated in ISL theories, which, due to their simplicity, have been widely used for polymer brushes. Such theories are based on Semenov's classical theory\cite{SST} that neglects fluctuations around the most probable chain trajectory and are strictly valid in the limit of infinite chain-stretching. Here we consider the ISL theory of Amoskov and Pryamitsyn\cite{Amos1+2,Amos3}, which is based on the Flory-Huggins theory\cite{FH} and for brushes of finitely extensible chains (e.g., lattice polymers) in an explicit solvent.\cite{Note2} To incorporate the correct solvent entropy in this theory, the right-hand-side of Eq.~(4) in Ref.~\cite{Amos3} needs to be multiplied by $r$. The predicted $\phi(x)$ of this modified theory is then shown in Figs.~\ref{Kent} and \ref{Karim}. We see that the agreement between the ISL and LSCF results becomes better for higher $\sigma$ and smaller $\chi$ (which make chains more stretched), and the differences between them are mainly due to the classical approximation and the infinite-stretching assumption used in the former (which are well understood in the case of continuous Gaussian chains\cite{Schk3}). We also note that in ISL theories all chains are grafted at $x=0$ with a reflecting substrate placed at $x=0$, unlike in our LSCF calculations; this leads to the difference in $\phi(x)$ near the substrate as shown in Figs.~\ref{Kent} and \ref{Karim}.

To summarize, we have showed how to properly account for the solvent entropy in polymer lattice models with multiple occupancy of lattice sites\cite{FLMC}, and presented a quantitative coarse-graining strategy that ensures both the solvent entropy and the fluctuations in the original systems are properly accounted for using such lattice models. Although proposed based on homogeneous polymer solutions, our strategy is equally applicable to inhomogeneous systems such as polymer brushes immersed in a small-molecule solvent.

This work was supported by NSF CAREER Award CBET-0847016.

\bibliographystyle{unsrt}

\end{document}